\documentclass[twocolumn,showpacs,preprintnumbers,amsmath,amssymb]{revtex4}

\usepackage{psfrag}
\usepackage{graphicx}
\usepackage{amsmath}
\usepackage{amsfonts}

\usepackage{latexsym}
\usepackage{epsfig}

\def\que{section}

\newtheorem{lemma}{Lemma}[\que]

\def\calC{{\mathcal C}}

\def\calH{{\mathcal H}}

\def\calO{{\mathcal O}}
\def\calP{{\mathcal P}}

\def\calS{{\mathcal S}}

\def\C{{\mathbb C}}   
\def\Z{\mathbb{Z}}    

\def\then{\Rightarrow}

\def\tr{{\rm tr}}              
\def\dim{\hbox{\rm dim}}       
\def\1{\hbox{1\hskip -3pt I}}  
\def\span{\hbox{\rm span}}     
\def\diag{{\rm diag}} 

\def\ovC{\overline{C}}

\def\ovr{\overline{r}}

\def\ovw{\overline{w}}

\def\un0{\underline{0}}

\def\bra#1{\,\langle{#1}|\,}                 
\def\ket#1{\,|{#1}\rangle\,}                 
\def\diada#1#2{\,|{#1}\rangle\langle{#2}|\,} 

\def\1{\openone}

\begin{document}

\preprint{Submitted to {\it Phys. Rev. A}}

\title{Pure entangled states probed by multipartite Bell experiments}

\author{Jes\'us Ur\'ias}
\affiliation{%
Instituto de F\'isica, UASLP\\
San Luis Potos\'i, SLP, M\'exico
}

\date{\today}

\begin{abstract}
The spectral decomposition of all $(n,2,2)$ Bell operators ($2^{2^n}$ in
number, $n \ge 2$), as introduced by Werner and Wolf~\cite{werner-01}, is done.
Its implications on the characterization of Bell operators as probes of
entanglement are considered in detail.
\end{abstract}


\maketitle

\section{\label{introduction}%
Introduction}

Bell experiments are designed to test correlations that are supported by local
realistic models~\cite{clauser-74}. A general experimental setup consists of
$n$ observers, each of them is given a particle and, independently, decide what
of $m$~observables, which are $v$-valued each, to measure.  All such setups
constitute the $(n,m,v)$ Bell experiment. One setup in the $(n,m,v)$ class may
be tuned up in one of $m^n$ forms and in each of its forms it detects one of
$v^n$ data instances. Bell inequalities are constraints that local realistic
models impose to the experimental correlation functions.

 The set of CHSH inequalities\cite{clauser-69} is complete for the $(2,2,2)$
class.  Mermin's example~\cite{mermin-90} generalizes CHSH inequalities to the
$(n,2,2)$ class, where a complete set of inequalities grows as
$2^{2^n}$~\cite{werner-01,zukowski} in cardinality. A complete theory of
$(n,2,2)$ Bell experiments was developed by Werner and Wolf
in~\cite{werner-01}.

 In the quantal theory the measurement in a Bell experiment corresponds to a
hermitian operator~$B$. For a state~$\psi$, the equivalent to Bell's inequality
is $|\tr(\psi B)| \le 1$. Every separable state satisfies the inequality.
Stronger quantal correlations may violate it.  When this happens for some
state~$\psi$ we say experiment $B$ probes entanglement.

We consider the infinite sequence of Bell operators for $(n,2,2)$ experiments,
the largest class of experimental setups for which a complete theory is
available~\cite{werner-01}. We assume the observables $A_k(0)$ and $A_k(1)$ per
particle $k\in\{1,\ldots, n\}$ with $A_k(s_k)^2 = \1$, $s_k=0$,~1.  In the
theory~\cite{werner-01}, for each $n$, every Bell operator, $B_f$, has index
$f$ that takes value in the set $\{-1,1\}^{2^n}$, of cardinality~$2^{2^n}$.
The set of $B_f$ operators is complete.

How a measurement $B_f$ must be in order to detect pure entangled states?
Answer to this question is important since the number of Bell inequalities
grows exceedingly fast with~$n$.  For the $(2,2,2)$ class, Gisin's
theorem~\cite{gisin-91} says that every pure nonproduct state of two particles
violates a CHSH~\cite{clauser-69} inequality.  However, the converse to Gisin's
theorem is not true. Actually, an experiment~$B_f$ does not detect every pure
entangled state. Several examples considered by Braunstein~{\it
et~al.}~\cite{braunstein-92} show that for detectability of pure entangled
states in a $(2,2,v)$ setup it is necessary that an eigenspace $\calH_\lambda$
of $B^2$ with a degenerate eigenvalue $\lambda^2 > 1$ exists. I.e,
$\dim(\calH_\lambda) > 1$ for some $\lambda > 1$.

A general technical lemma (Lemma~\ref{proposition}) lead us to the
characterization of a $(n,2,2)$ Bell experiment by the following equivalent
assertions.

\medskip

 (1.a) $B_f$ is a probe of entangled states.

 (1.b) $B_f$ has spectral radius greater than one.

 (1.c) There exists an eigenspace $\calH_\lambda$ of $B_f^2$ with

  $\dim(\calH_\lambda) > 1$ for some $\lambda > 1$.

\medskip

We are expressing characteristics of $B_f$ in terms of those ones of $B_f^2$
because in Section~\ref{experiments-section} we show that $B_f^2$ has a simple
spectral decomposition and the set of eigenstates constitutes a ``natural''
product basis for the state space,~$\calH$.
Main results on $B_f^2$ are proved
in Appendix~\ref{appendix-coefficients}. In particular, formulae are given to
compute the spectral radius, $\Lambda_f$, of $B_f$ in terms of the geometric
parameters of the experimental setup, for each~$f$. The spectral radius
$\Lambda_f$ is the largest factor of violation that is detectable by experiment
$B_f$.  The maximal violation factor that is allowed in principle for $(n,2,2)$
Bell experiments was determined by Werner and Wolf in~\cite{werner-01} to be
$\Lambda_n \equiv 2^{(n-1)/2}$. I.e., $\Lambda_f \le \Lambda_n$ for every $f$.
The equality holds, e.g., for Mermin's example.

Once we know that $B_f$ detects entanglement, by a violation factor of at most
$\Lambda_f$, we want a finer characterization of it.  
  What entangled states are detectable by $B_f$? 
  What $f$ and what geometry make experiment $B_f$ to admit the maximal
  violation factor $\Lambda_n$ that is allowed {\it in principle}?

The spectrum $\lambda_f^2$ of $B_f^2$ satisfies the sum rule
\begin{equation}\label{sum-rule}
  \sum_{w} \lambda_f^2(w) = \dim(\calH) = 2^n
\ ,
\end{equation}
where $w$ is an index for the eigenstates of $B_f^2$. The total
spectrum~(\ref{sum-rule}) is independent of $f$ and of geometric parameters.
Thus, the difference between experiments is how the total spectrum is
distributed among states. In extreme situations are an experiment with spectrum
equally distributed among all states and an experiment with spectrum
concentrated in just two states. In the first case the sum
rule~(\ref{sum-rule}) implies that $\lambda_f(w) = 1$ and the experiment does
not detect entanglement. In the second case sum rule~(\ref{sum-rule}) implies
that the maximal violation factor~$\Lambda_n$ is admitted by the experiment.
Thus, we say that~$f$ is optimal if $B_f$ has a geometry where $\Lambda_f =
\Lambda_n$.  An experiment $B_f$ is ``optimal'' in the sense that it
``concentrates'' the maximal factor of violation $\Lambda_n$ in as few
entangled states as possible.  Such an experiment is the sharpest test for
entanglement.  The following equivalent assertions provide a characterization
of optimal $(n,2,2)$~Bell experiments.

\medskip

(2.a) $B_f$ has spectral radius $\Lambda_n$

(2.b) $f$ is optimal and $\|(i/2)[A_k(0),A_k(1)]\| = 1$ for

      each particle $k$.

(2.c) The Bell inequality is violated by two and only

      two eigenstates of $B_f$.

\medskip

We find that, {\it for each $n$, there are four optimal $(n,2,2)$ Bell
experiments of which only two are independent}. A very small number as compared
with $2^{2^n}$, even for small~$n$. And only two of their eigenstates violate
Bell's inequality. I.e., (2.c)~above.

A formula to compute the
optimal values of $f$ for each $n$ is given.  Optimal probes $B_f$ are computed
easily. We provide examples for $n=2$, 3 and 4 in
Appendix~\ref{appendix-examples}. For $n=2$ they are the CHSH
operators~\cite{clauser-69} and for $n=3$ the Bell polynomials were derived
in~\cite{werner-01}.

The entangled character of the states detected by an optimal experiment depends
on the geometry.  Statement~(2.b) has room for $2^n$ geometries.  {\it For any
$f$, all eigenstates of $B_f$ are generalized GHZ states}: superposition of two
states with antipodal configurations. For optimal $f$ the sum
rule~(\ref{sum-rule}) is saturated by just a pair of antipodal configurations.
Every geometry determines a pair and there two geometries per pair. For
example, in a CHSH setup there are four geometries that admit the maximal
violation factor: two geometries ``put'' the violation factor on two EPR states
and two other geometries put it on two GHZ states.

\section{\label{experiments-section}
$(n,2,2)$ Bell operators}

We consider the infinite sequence of Bell operators introduced by Werner and
Wolf for which the theory is developed in~\cite{werner-01} to be as explicite
and complete as in the CHSH case.~\cite{clauser-69} 
To make the article self-contained, the next paragraph summarizes the main line
of thought in~\cite{werner-01}.

Each $(n,2,2)$ Bell experiment admits a set $\calS := \{0,1\}^n$ of
experimental setups. It is convenient to consider $\calS \equiv \Z_2^n$. For
every setup $s \in \calS$ there is the correlation function $\xi(s) =
\left\langle \prod_{k=1}^n A_k(s_k) \right\rangle$ which is considered to be
the $s$-coordinate of vector~$\xi$. The local-realistic hypothesis bounds
vector $\xi$ to lay in the convex hull $\Omega$ of the finite collection of
vectors $\{\pm\epsilon_r:\ r\in \{0,1\}^n \}$, with coordinates $\epsilon_r(s)
= (-1)^{\langle r, s \rangle}$. The opposite is also true. This assertion is
equivalent to say that $\langle \beta, \xi \rangle \le 1$ for each vector
$\beta \in \Omega^\circ$, the polar set of $\Omega$. Being $\Omega^\circ$ a
polytope it is enough to ask that $\langle \hat{f}, \xi \rangle \le 1$ for each
maximal vector $\hat{f}$ of $\Omega^\circ$. Maximal vectors are found to be
given by the Fourier transform on the group $\calS$
\begin{equation}\label{fourier-transform}
  \hat{f}(s) = 2^{-n} \sum_{r \in \calS} (-1)^{\langle r,s \rangle} f(r)
\ ,
\end{equation}
for each $f: \calS \to \{-1,1\}$.

In the quantal description the quantities $\langle \hat{f},\xi\rangle = \sum_{s
\in \calS} \hat{f}(s) \left\langle \prod_{k=1}^n A_k(s_k) \right\rangle$ are
replaced by the expectation value of the operators
\begin{equation}\label{bell-operator}
  B_f = \sum_{s \in \calS} \hat{f}(s) \bigotimes_{k=1}^n A_k(s_k)
\ ,
\end{equation}
one for each $f \in \{-1,1\}^{2^n}$. The freedom we have to choose observables
makes $B_f$ to depend on geometric parameters. Thus, every vector $f$
represents, not one, but a class of Bell operators.

In Appendix~\ref{appendix-coefficients} we show that the operator $B_f^2$ is
given by the formula
\begin{equation}\label{b-squared}
 B_f^2
 = \1 + \sum_{\substack{p \subset \{1,\ldots,n\}  \\ \# p~\text{is
   even}}}  C_p(f) \, \bigotimes_{k \in p} \frac{i}{2} [A_k(0),\,A_k(1)]
\end{equation}
where the sum is over all non-empty subsets of points $p \subset \{1, \ldots, n
\}$ of even cardinality.  Coefficients $C_p(f)$ depend on geometric parameters
and they are $2^{n-1}-1$ in number. Everyone is bounded to lay in the interval
$[-1,1]$,
\begin{equation}\label{c-bounded-to-one}
                      |C_p(f)| \le 1
\ .
\end{equation}
Formulae to calculate $C_p(f)$ are given in
Appendix~\ref{appendix-coefficients}.  Equality in~(\ref{c-bounded-to-one})
holds for a vector $f$ in Mermin's example.

The spectral decomposition of Bell operator $B_f$ in
Section~\ref{detectability-section} follows from the simple spectral properties
of operator $B_f^2$ in~(\ref{b-squared}). For the moment, remark that each
operator $(i/2)[A_k(0), A_k(1)]$ in~(\ref{b-squared}) is hermitian and laying
in the ball $\|A\| \le 1$. Thus, it can be represented in the form
\begin{equation}\label{z-operator}
 \frac{i}{2} [A_k(0),\,A_k(1)] = \sin\theta_k Z_k,
 \quad \quad \theta_k \in [-\pi/2,\,\pi/2]
 \ ,
\end{equation}
where each hermitian operator $Z_k$ is traceless and maximal, $\|Z_k\| = 1$.
The obvious choice is to take $Z_k = \sigma_3$, the same at each site~$k$.
It amounts to take $A_k(s_k) = \langle n_k(s_k), \sigma \rangle$ with unitary
vector $n_k(s_k) = (\cos\varphi_k(s_k)$, $\sin\varphi_k(s_k)$, $0)$ laying on
the $x$-$y$--plane of a local coordinate system such that $n_k(1) \times n_k(0)
= (0$, 0, $\sin\theta_k)$.

  The orthonormal set $\{\ket{w_k}:\ w_k= -1, 1\}$ of eigenvectors of
$\sigma_3$  is adopted as the basis of the state space $\calH_k = \C^2$.  The
set of configurations $\calC$ for the product basis of $\calH = \C^{2 \otimes
n}$ is $\calC = \{-1, 1\}^n$.  For each configuration $w = w_1 w_2 \cdots w_n
\in \calC$ there is the product vector
\begin{equation}\label{product-vector}
 \ket{w} = \ket{w_1} \otimes \ket{w_2} \cdots \otimes \ket{w_n}
\end{equation}
which is an element of the product basis of $\calH$. Thus, each product
vector~(\ref{product-vector}) is an eigenvector of $B_f^2$ with eigenvalue
\begin{equation}\label{b2-eigenvalues}
 \lambda_f^2(w)
 = 1 + \sum_{\substack{p \subset \{1, \ldots, n\} \\ \# p~\text{is even}}}
   C_p(f) \prod_{k \in p} w_k \sin\theta_k \quad \ge \quad 0
\ .
\end{equation}

The spectral radius of $B_f$ is $\Lambda_f := \max_w\{\lambda_f(w)\}$. It is
attained by a configuration $W \in \calC$ such that $C_p(f) \prod_{k \in p}
W_k \sin\theta_k \ge 0$. I.e., $\Lambda_f = \lambda_f(W)$. The spectral radius
may be computed with the formula
\begin{equation}\label{spectral-radius}
  \Lambda_f
  =
  1 + \sum_{\substack{p \subset \{1, \ldots, n\} \\ \# p~\text{is even}}}
  |C_p(f)| \prod_{k \in p} |\sin\theta_k|
\end{equation}
and results on $C_p(f)$ in Appendix~\ref{appendix-coefficients}.

The following symmetries of $B_f^2$ and its eigenvalues are apparent
from~(\ref{b-squared}) and~(\ref{b2-eigenvalues}).

\medskip

\noindent (3.a)
  $B_f^2$ is invariant under the exchange of observables $A_k(0)
\leftrightarrow A_k(1)$ at all points~$k$. I.e., $\sin\theta_k \leftrightarrow
-\sin\theta_k$ and there are at least two geometries that yield identical
results.

\medskip

\noindent (3.b)
  $\lambda_f(w) = \lambda_f(\tilde{w})$, where the configuration $\tilde{w} \in
\calC$ has coordinates $\tilde{w}_k = - w_k$. I.e., the eigenvalues of $B_f^2$
are at least doubly degenerate.

\medskip

Due to the form~(\ref{b-squared}) of $B_f^2$, the
eigenvalues~(\ref{b2-eigenvalues}) satisfy sum rule~(\ref{sum-rule}).
The value for the sum~(\ref{sum-rule}) is independent of any geometric
parameter and of $f$.

Results~(\ref{b2-eigenvalues}) and~(\ref{sum-rule}) allow us to think of the
spectral function $\lambda^2_f$ as a weight function on the set of
configurations~$\calC$ with full weight $\#\calC = 2^{n}$. The same amount of
total weight is available to every experiment $B_f$, independent of $f$ and of
the choice of observables.  The difference between experiments consists in the
way they distribute the weight in the configuration set $\calC$. An experiment
that equally distributes the weight among all configurations does not probe
entanglement since $\lambda_f^2(w) = 1$ for each $w \in \calC$. This happens
when $[A_k(0), A_k(1)] = 0$ at every point~$k$. 

Experiment $B_f$ is a good probe of entanglement if the total weight $\#\calC$
is supported in as few configurations as possible. In
Section~\ref{maximal-subsection} we will see that the ``best'' situation is
when $\lambda_f^2$ concentrates all of the weight that is available, $\#\calC$,
on just two configurations.

\section{\label{detectability-section}%
                    All Bell eigenstates are GHZ}
The spectral decomposition of $B_f^2$ is simple enough as to base on it the
corresponding decomposition of $B_f$. To proceed, remark that $B_f^2$ is a
positive operator and it has a unique positive square root, denoted by $|B_f|$.
Operator $|B_f|$ has the same set $\{\ket{w}\}$ of eigenvectors as $B_f^2$,
with eigenvalues $\lambda_f(w) \ge 0$.  Experiment~$B_f$ is decomposed into the
product $B_f = S_f\,|B_f|$ where, in general, the operator $S_f$ is an
isometry. For the particular form of operator $B_f$ in~(\ref{bell-operator}) a
direct calculation shows that $B_f \ket{w} = \beta_f(w) \ket{\tilde{w}}$, with
$|\beta_f(w)| = \lambda_f(w)$.  Thus, up to a phase factor, $S_f$ is the
permutation,
\begin{equation}\label{permutation}
  S_f \ket{w} = \ket{\ovw} := e^{i\varphi_f(w)} \ket{\tilde{w}}
\ ,
\end{equation}
where the phase factor, whenever $\lambda_f(w) \ne 0$, can be written as
\begin{equation}\label{phase-factor}
  e^{i\varphi_f(w)} = \frac{\beta_f(w)}{\lambda_f(w)}
\ .
\end{equation}
Furthermore, $S_f^2 = \1$. 
The properties of $S_f$ are used in the following
construction.

Let $\calH_\lambda \subset \calH$ be an eigenspace of $B_f^2$ corresponding to
the eigenvalue $\lambda^2$. Consider product state $\ket{w} \in \calH_\lambda$.
By~(\ref{permutation}), the vectors
$$
  B_f\ket{w} = \lambda \ket{\ovw}
\quad\text{and}\quad
  B_f\ket{\ovw} = \lambda \ket{w}
$$
are orthonormal, in $\calH_\lambda$ too. It is of course necessary that
$\dim(\calH_\lambda) > 1$.  That it is so, follows from the properties of
$S_f$. Then, the 2-$d$ subspace 
$
\span\{\ket{w}, \ket{\tilde{w}}\} \subset \calH_\lambda
$
is $S_f$-invariant and every eigenvector of all Bell operators $B_f$ is an
entangled state of the form
\begin{equation}\label{b-eigenvectors}
  \ket{w; \pm}
  =
  \frac{1}{\sqrt{2}}
  \left( \ket{w} \pm \ket{\ovw} \right)
\end{equation}
with eigenvalues $\pm\lambda = \pm \lambda_f(w)$. Any $(n,2,2)$ Bell experiment
probes entanglement of states with antipodal configurations, $w$ and
$\tilde{w}$. Such pure states we call GHZ.

Remark that two GHZ states $\ket{w;\pm}$ and $\ket{\tilde{w};\pm}$ differ by
just a phase factor. Thus, to deal with the eigenvectors of $B_f$ we consider
configurations in the quotient set $\calC/{\hskip - 0.4ex \sim}$, where
antipodal configurations $w$ and $\tilde{w}$ are equivalent.

That $\dim(\calH_\lambda) > 1$ is a necessary and sufficient condition for the
existance of pure entangled states that violate Bell inequality by the factor
$\lambda > 1$ is proved in general in the following.

{\lemma\label{proposition}
 Let $\calH_\lambda$ be an eigenspace of $B^2$ of eigenvalue $\lambda^2$.
Assume $\lambda > 1$. Then {\rm (I)}~there exists $\ket{\,} \in \calH_\lambda$
such that
\begin{equation}\label{si-viola}
|\bra{\,}B\ket{\,}| \  > \  1
\end{equation}
if and only if {\rm (II)}~$\dim(\calH_\lambda) \ge 2$.
}

\noindent{\bf Proof. }
(II $\then$ I). We are assuming that $\lambda^2 > 1$ is a degenerate eigenvalue
of $B^2$. The eigenspace $\calH_\lambda$ of $B^2$ is $B$-invariant. The
restriction $B|_{\calH_\lambda}$ is hermitian and thus it has eigenvectors
$\ket{\lambda} \in \calH_\lambda$ with eigenvalues $\lambda$ such that
$|\lambda| > 1$ Then, statement~(I) follows. Remark that
$\ket{\lambda}$ is not separable by necessity. Such eigenvectors were
actually constructed in~(\ref{b-eigenvectors}) for the particular form of the
$(n,2,2)$ $B_f$ operators.

Sufficiency is proved by contradiction ($\neg$II $\then$ $\neg$I). 

\noindent
We are assuming $\lambda > 1$ and by
hypothesis~($\neg$II) the eigenspace $\calH_\lambda$ of $B^2$ has
$\dim(\calH_\lambda) = 1$ (the case of zero dimension is trivial) and is
spanned by the eigenvector $\ket{\lambda}$. Since $\calH_\lambda$ is
$B$-invariant then $\ket{\lambda}$ is an eigenvector of $B$ too. The
contradiction stems from the fact that the unique state $\ket{\lambda} \in
\calH_\lambda$ is a product vector~(\ref{product-vector}) for some
configuration $w \in \calC$ and then inequality~(\ref{si-viola}) is violated by
the only (separable) state $\ket{w} \in \calH_\lambda$.  \hfill$\Box$

\medskip

In Section~\ref{experiments-section} we proved that each eigenvalue of
$B_f^2$ is at least doubly degenerate. Thus Lemma~\ref{proposition} applies and
a corollary is the following characterization of $(n,2,2)$ Bell experiments as
probes of entanglement: {\it $B_f$ detects pure entangled states iff $B_f^2$
has a spectral radius that is greater than one}.

Eigenvectors of $B_f$ come in entangled pairs of the
GHZ~form~(\ref{b-eigenvectors}), two such states for each $w \in \calC/{\hskip
-0.4ex \sim}$.  In the GHZ basis {\it all} $(n,2,2)$ Bell operators have the
diagonal form $B_f = \diag(B_f(+), B_f(-))$ with diagonals 
$$
  B_f(\pm) = \pm \sum_{w \in \calC/{\hskip -0.4ex
\sim}}\lambda_f(w)\diada{w;\pm}{w;\pm} 
\ .
$$

Every entangled state that is detectable by experiment $B_f$ is supported on
the subspace $\calH_> := \bigoplus_{\lambda > 1} \calH_\lambda$.  The factors
of violation associated with the entangled state~(\ref{b-eigenvectors}) is
$\lambda_f(w)$.  The definition of state $\ket{\ovw}$ includes the phase
factor~(\ref{phase-factor}) that we did not calculate.

\section{\label{maximal-subsection}%
                         Optimal probes}
The upper bound $\Lambda_n = 2^{(n-1)/2}$ for the maximal possible violation
factor of any $(n,2,2)$ Bell inequality was determined by Werner and Wolf
in~\cite{werner-01}. Here, directly from~(\ref{b-squared})
and~(\ref{c-bounded-to-one}), we see that
\begin{eqnarray}\label{upper-bound}
 \Lambda_f
 &\le&
 \max_f \|B_f^2\|^{1/2}
 \le \left( 1 + \sum_{m=1}^{\lfloor n/2 \rfloor}
     \binom{n}{2m} \right)^{1/2}
\nonumber \\
 &\le& 2^{(n-1)/2} \ .
\end{eqnarray}
The bound is attained by an experiment $B_f$ with a configuration~$w$ and
geometry such that 
$$
   |w_k \sin\theta_k| = \|\frac{i}{2}[A_k(0),A_k(1)]\| = 1.
$$
and a vector $f \in \{-1,1\}^{2^n}$ such that $\ovC_p(f) = 1$, where
$\ovC_p(f)$ is the value of $C_p(f)$ when geometric parameters are
$\cos\theta_k = 0$. A formula for $\ovC_p(f)$ is given
in Appendix~\ref{appendix-examples}, formula~(\ref{cp-perpendicular}).

Under such conditions, formula~(\ref{b2-eigenvalues}) gives $\lambda_f^2(w) =
2^{n-1}$ as the largest eigenvalue possible for the given $B_f^2$.  Thus, {\it
the $(n,2,2)$~Bell experiment $B_f$ has maximal spectral radius  $\Lambda_n
= 2^{(n-1)/2}$ if and only if $\ovC_p(f) = 1$ and $\|(i/2)[A_k(0),A_k(1)]\| = 1
$}.

Let us assume experiment $B_f$ admits the maximal violation factor $\Lambda_n =
2^{(n-1)/2}$. Then, according to Lemma~\ref{proposition}, the corresponding
Bell inequality is violated iff there exists at least two eigenstates $\ket{w}$
and $\ket{\ovw}$ of $B_f^2$ with eigenvalue $\lambda_f^2(w) = \lambda_f^2(\ovw)
= 2^{n-1}$.  But two such states saturate the sum rule~(\ref{sum-rule}) and
there can not be any other eigenstate involved in the violation of Bell
inequality (just one state is not enough because eigenstates of $B^2$ are
separable).  Thus, in experiments of maximal violation, the spectrum
$\lambda_f^2$ equally distributes weight $\#\calC$ in just two configurations
(no less, no more).  This proves that {\it $B_f$ has spectral radius
$\Lambda_n$ if two and only two of its eigenstates violate the corresponding
Bell inequality}.

Experiments of maximal violation are optimal probes of entanglement in that
they are tunable as to detect any given configuration $w$ by choosing the sign
of $\sin\theta_k = \pm 1$ such that $w_k\sin\theta_k = 1$.  In an optimal Bell
experiment $\lambda_f^2$ concentrates all weight that is available in just two
GHZ states of our choice: $\ket{w;+}$ and $\ket{w;-}$.

Looking at formula~(\ref{cp-perpendicular}) for $\ovC_p(f)$ we immediately see
that a vector $f$ that satisfies the conditions
\begin{equation}\label{optimal-f}
  f(s)f(s+p) = (-1)^{\langle p,s \rangle + \#p/2},
\end{equation}
is an optimal one with $\ovC_p(f) = 1$. In~(\ref{optimal-f}), $s \in \calS$ and
the subset $p\subset\{1,\ldots,n\}$, of even cardinality, is represented by the
vector $p\in\calS$ with coordinate $p_k=1$ iff $k \in p$. Remark that for any
two subsets $p$ and $p'$ of even cardinality, the symmetric difference $p+p'$
(coordinate-wise mod~2 addition) is also of even cardinality. The collection
$\calP_n := \{p:\ \#p =$~even$\}$ (when $0\cdots0$ is included) is a subgroup
of $\calS$ of order $2^{n -1}$. So the set of independent
conditions~(\ref{optimal-f}) lay on the orbits of two setups. E.g., $s=
0\cdots00$ and $s = 0\cdots01$. The sign of the coordinates of $f$ may be
assigned in just two ways to each orbit. Thus, {\it for each $n$, there are four
optimal $(n,2,2)$ Bell experiments of which only two are independent}. 

Optimal probes of entanglement for any number $n$ of particles are computed
easily by using formula~(\ref{optimal-f}). Probes for $n=2$ (the CHSH
operators\cite{clauser-69}), $n=3$ (derived in~\cite{werner-01}) and~$n=4$ are
obtained in Appendix~\ref{appendix-examples}.

\appendix 
\section{\label{appendix-coefficients}%
                       The operator $B_f^2$ }
Results~(\ref{b-squared}) and~(\ref{c-bounded-to-one})
about $B_f^2$ are obtained in the following. 

From definition~(\ref{bell-operator}) we have that
\begin{eqnarray}
  B_f^2
  &=& \sum_{s,s' \in \calS} \hat{f}(s)\hat{f}(s') \bigotimes_{k=1}^n
  A_k(s_k) A_k(s_k')
\label{b-squared-app} \\
  &=& \sum_{s,s' \in \calS} \hat{f}(s)\hat{f}(s') \calO(s,s')
\nonumber
\end{eqnarray}
where
\begin{equation}\label{o-symmetric}
  \calO(s,s')
  = \text{Sym}_{s,s'} \bigotimes_{k=1}^n
  \left( \Phi_k(s_k,s_k') + i \Gamma_k(s_k,s_k') \right)
\end{equation}
is the symmetric component of the tensor-product operator
in~(\ref{b-squared-app}).
I.e., $\calO(s,s') = \calO(s',s)$.  In~(\ref{o-symmetric}) we have made use of
the following definitions
\begin{eqnarray}\label{phi}
  \Phi_k(s_k,s_k')
  &:=& \frac{1}{2}\{A_k(s_k), A_k(s_k')\}
\nonumber \\
  &=& \openone \frac{ 1 + (-1)^{s_k + s_k'} }{2}
\\
  & &  + \frac{1}{2}\{A_k(0), A_k(1)\} \frac{1 - (-1)^{s_k + s_k'}}{2}
\nonumber
\end{eqnarray}
and
\begin{eqnarray}
 \Gamma_k(s_k,s_k')
 &:=& \frac{-i}{2}[A_k(s_k), A_k(s_k')] 
\\
 &=& \frac{(-1)^{s_k} - (-1)^{s_k'}}{2}
     \,\,\frac{i}{2}[A_k(1), A_k(0)]
\ , \nonumber
\end{eqnarray}
The hermitian operators $\Phi_k(s_k, s_k')$ and $\Gamma_k(s_k,s_k')$ are
symmetric and anti-symmetric, respectively, under the exchange $s_k
\leftrightarrow s_k'$. Furthermore, under the assumption that $A_k(s_k)^2 =
\1$, we have that
 $$[\Phi_k(s_k, s_k'), \Gamma_k(s_k,s_k')] = 0 \ .$$
These properties allow us to write the symmetric
operator~(\ref{o-symmetric}) in the following form,
\begin{equation}\label{o-in-evens}
  \calO(s,s')
  = \sum_{\substack{p \subset \{1,\ldots,n\}  \\ \# p~\text{is
   even}}} \bigotimes_{k \in p^c} \Phi_k(s_k,s_k') \bigotimes_{k \in p}
  \Gamma_k(s_k, s_k')
\ ,
\end{equation}
where the sum is over subsets $p \subset \{0, \ldots, n\}$ of even
cardinality, including $p = \emptyset$.

Substituting~(\ref{o-in-evens}) in~(\ref{b-squared-app}) we obtain
\begin{eqnarray}
  B_f^2
 &=& \1 + \sum_{\substack{p \subset \{1,\ldots,n\}  \\ \# p~\text{is
   even}}} (-1)^{\#p/2} \sum_{s,s' \in \calS} \hat{f}(s)\hat{f}(s')
\nonumber \\
 & & \hskip 6 ex
 \bigotimes_{k \in p}\Gamma_k(s_k,s_k') \bigotimes_{\ell \in p^c}
 \Phi_\ell(s_\ell,s_\ell')
 \label{b-squared-after-o}
\end{eqnarray}
where the term~$\1$ is the contribution from $p = \emptyset$ and the sum now
runs over non-empty subsets.  Considering in~(\ref{b-squared-after-o}) that
$(1/2)\{A_k(0), A_k(1)\} = a_k \1$ with $a_k = \cos\theta_k$, we
get formula~(\ref{b-squared}) for $B_f^2$ with coefficients, at the moment,
given by
\begin{equation}\label{coeff-app}
 C_p(f) = (-1)^{\#p/2} 2^{-n} F_f(p)
\end{equation}
where
\begin{equation}
  F_f(p) := \sum_{s,s' \in \calS} \hat{f}(s)\hat{f}(s')
  \prod_{k \in p} \gamma(s_k,s_k')\prod_{\ell \in p^c} \varphi_k(s_k,s_k')
\ , \label{coefficients-in-b}
\end{equation}
\begin{equation}\label{n-gamma}
 \gamma(s_k,s_k') = (-1)^{s_k} - (-1)^{s_k'}
\end{equation}
and
\begin{equation}\label{n-phi}
 \varphi_k(s_k,s_k') = 1 + (-1)^{s_k + s_k'} + a_k (1 - (-1)^{s_k + s_k'})
\ .
\end{equation}
The products in~(\ref{coefficients-in-b}), using~(\ref{n-gamma})
and~(\ref{n-phi}), expand to
\begin{equation}\label{product-gamma}
 \prod_{k \in p} \gamma(s_k,s_k')
 = \sum_{r \subset p} (-1)^{\#r} (-1)^{\sum_{k \in r}s_k' + \sum_{k \in r^c}
   s_k} \ ,
\end{equation}
where $r \cup r^c = p$, and to
\begin{eqnarray}\label{product-phi}
 \prod_{k \in p^c}\varphi_k(s_k,s_k')
 &=& \sum_{q \subset p^c}(-1)^{\sum_{k \in q^c}(s_k + s_k')} 
     \prod_{k \in q}(1+a_k)
\nonumber \\
  & & \hskip 11 ex \prod_{k \in q^c} (1-a_k)
\ ,
\end{eqnarray}
where $q \cup q^c = p^c$. The result for $F_f(p)$ in~(\ref{coefficients-in-b})
with the products expanded is
\begin{equation}\label{f-j}
  F_f(p) = \sum_{q \subset p^c} \prod_{k \in q} (1 + a_k) \prod_{\ell \in q^c}
  (1-a_k) \sum_{r \subset p} (-1)^{\# r} G(q,r)
\end{equation}
where
\begin{equation}\label{two-ifts}
  G(q,r) = \sum_{s,s' \in \calS} \hat{f}(s)\hat{f}(s') (-1)^{\sum_{k \in (q
  \cup r^c)}s_k + \sum_{k \in (q \cup r)}s_k'}
\ .
\end{equation}
For the following, it is convenient to denote subsets $q \subset \{1, \ldots,
n\}$ by vectors $q \in \calS$ such that $k \in q$ iff $q_k = 1$ (we are abusing
notation but no confusion will arise). With this identification, $G(q,r)$
in~(\ref{two-ifts}) is seen to be the product of two inverse Fourier transforms,
\begin{equation}\label{simple-g}
  G(q,r) = f(q+\ovr) f(q+r) \ ,
\end{equation}
where $q\cup r \leftrightarrow q+r$ (because $qr = 0$) and $r + \ovr = p$.
Formula~(\ref{simple-g}) shows us that
\begin{equation}\label{g-one}
 |G(q,r)| = 1\ .
\end{equation}

We have everything to prove inequality~(\ref{c-bounded-to-one}).
From~(\ref{coeff-app}) and~(\ref{f-j}) we have that
$$
  |C_p(f)|
  \le
  2^{-\#p^c}
  \sum_{q \subset p^c} \prod_{k \in q} (1 + a_k) \prod_{\ell \in q^c} (1-a_k)
$$
where we have made use of~(\ref{g-one}) and the fact that $|a_k| \le 1$. One
proves by induction in $\#p^c$ that
\begin{equation}\label{induction}
\sum_{q \subset p^c} \prod_{k \in q} (1 + a_k) \prod_{\ell \in q^c} (1-a_k)
= 2^{\#p^c}
\end{equation}
and the inequality~(\ref{c-bounded-to-one}) follows. Our final answer for
$C_p(f)$ consists of formulae~(\ref{coeff-app}), (\ref{f-j}) and
(\ref{simple-g}).

\section{\label{appendix-examples}%
                          Examples}
The geometric parameters in optimal probes are settled down to $a_k = 0$. I.e.,
$\sin\theta_k = \pm1$.  Using the results obtained in
Appendix~\ref{appendix-coefficients}, the value $\ovC_p(f)$ of the
coefficient~(\ref{coeff-app}) for $a_k = 0$ is
\begin{equation}\label{cp-perpendicular}
  \ovC_p(f)
  =
  (-1)^{\#p/2} 2^{-n} \sum_{s \in \calS} (-1)^{\langle p,s \rangle}
  f(s) f(s+p)
\ ,
\end{equation}
for each non-empty subset $p \subset \{1, \ldots, n\}$ of even cardinality.
A vector $f$ that satisfies conditions~(\ref{optimal-f}) makes $\ovC_p(f) = 1$
and corresponds to an optimal experiment $B_f$. Optimal vectors for $n=2$, 3
and~4 are computed in the following.

For $n=2$ we have that $\calP_2 = \{11\}$, that corresponds to the only subset
$p = \{1,2\}$. The orbits of setups $00$ and $01$ for
conditions(\ref{optimal-f}) are
\begin{equation}\label{optimal-n=2}
 f(00)= - f(11)  \quad\text{and}\quad f(01) = (10)
\ .
\end{equation}
For instance the vector $f=(1$, 1, 1, $-1)$, with Fourier
transform $\hat{f}=(1/2$, 1/2, 1/2, $-1/2)$, is optimal and corresponds to one
of the CHSH operators.

For $n=3$ we have the collection $\calP_3 = \{011, 101, 110 \}$ that
produce the two orbits
\begin{eqnarray}
 f(000)=-f(011) = -f(101) = -f(110) \ , & &
 \nonumber \\
 f(001)= f(010) =  f(100) = -f(111) \ .& &
 \label{optimal-n=3}
\end{eqnarray}
Two independent choices of sign in~(\ref{optimal-n=3}) give the vectors
\begin{eqnarray}
 f_1 &=& (1,\ 1,\ 1,\ -1,\ 1,\ -1,\ -1,\ -1),
\nonumber\\
 f_2 &=& (1,\ -1,\ -1,\ -1,\ -1,\ -1,\ -1,\ 1),
\label{f-n=3}
\end{eqnarray}
with Fourier transforms
\begin{eqnarray}
 \hat{f}_1 &=& \frac{1}{2}(0,\ 1,\ 1,\ 0,\ 1,\ 0,\ 0,\ -1)
\ ,\nonumber\\
 \hat{f}_2 &=& \frac{1}{2}(-1,\ 0,\ 0,\ 1,\ 0,\ 1,\ 1,\ 0)\label{ft-n=3}
\end{eqnarray}
and Bell operators
\begin{eqnarray*}
  B_{f_1} &=& \frac{1}{2} \big( A_1(0)A_2(0)A_3(1) +  A_1(0)A_2(1)A_3(0)
\\
      & &  + A_1(1)A_2(0)A_3(0) -  A_1(1)A_2(1)A_3(1) \big) 
\ , \\
  B_{f_2} &=& \frac{1}{2} \big( -A_1(0)A_2(0)A_3(0) +  A_1(0)A_2(1)A_3(1)
 \\
      & &  + A_1(1)A_2(0)A_3(1) +  A_1(1)A_2(1)A_3(0) \big) 
\ .
\end{eqnarray*}
Luckily(?), for $n=3$ half of the coordinates of vectors $\hat{f}$ in
(\ref{ft-n=3}) vanish. Thus, half of the experimental setups are not involved
in the optimal operators~$B_f$. Same situation happens for $n=5$ particles.

Last example is for $n=4$ with collection of even vectors $\calP_4 =\{1100,
1010, 1001, 0110, 0101, 0011, 1111\}$ that produces the orbits
$$
\begin{cases}
 f(0000) = -f(0011) = -f(0101) = -f(0110)\\
 = -f(1001) = -f(1010) = -f(1100) = f(1111)
\end{cases}
$$
and
$$
\begin{cases}
 f(0001) = f(0010) = f(0100) = -f(0111)\\
 = f(1000) = -f(1011) = -f(1101) = -f(1110)
\end{cases}
$$
The choice of sign $f(0000) = f(0001) = 1$ in the orbits gives the optimal
vector
$$
 f= (1,1,1,-1,1,-1,-1,-1,1,-1,-1,-1,-1,-1,-1,1)
$$
with Fourier transform
\begin{equation}\label{optmial-n=4}
 \hat{f} = \frac{1}{4}(1,-1,-1,-1,-1,-1,-1,1,-1,-1,-1,1,-1,1,1,1)
\end{equation}
and a Bell operator with sixteen products of four observables each. The reader
may construct $B_f$ himself by using~(\ref{bell-operator}) with the optimal
vector~(\ref{optmial-n=4}).


\begin{thebibliography}{33}
\bibitem{braunstein-92}
 S.L. Braunstein, A. Mann and M. Revzen,
 Maximal violation of Bell inequalities for mixed states.
 {\it Phys. Rev. Lett. \bf 68} (1992) 3259--3261.
\bibitem{clauser-69}
  J.F. Clauser, M.A. Horne, A. Shimony and R.A. Holt,
  Proposed experiment to test local hidden-variable theories.
  {\it Phys. Rev. Lett. \bf 23} (1969) 880--884.
\bibitem{clauser-74}
 John F. Clauser and Michael A. Horne,
 Experimental consequences of objective local theories,
 {\it Phys. Rev. D. \bf 10} (1974) 526--535.
\bibitem{mermin-90}
 N. David Mermin,
 Extreme quantum entanglement in a superposition of macroscopically distinct
 states.
 {\it Phys. Rev. Lett. \bf 65} (1990) 1838--1840.
\bibitem{gisin-91}
 N. Gisin,
 Bell's inequality holds for all non-product states.
 Phys. Lett. A 154 (1991) 201--202.
\bibitem{werner-00}
 R.F. Werner and M.M. Wolf, Bell's inequalities for states with positive
partial transpose. {\it Phys. Rev. A \bf 61} (2000) 062102.
\bibitem{werner-01}
 R.F. Werner and M.M. Wolf, All multipartite Bell correlation inequalities for
 two dichotomic observables per site. 
 {\it arXiv}:quant-ph/0102024 v1 (2001).
\bibitem{zukowski}
 M. Zukowski and C. Brukner,
 Bell's theorem for general $N$-qubit states.
 {\it arXiv}:quant-ph/0102039 v3 (2002).
\end{thebibliography}
\end{document}